\documentclass{article}

\usepackage{arxiv}

\usepackage[utf8]{inputenc} 
\usepackage[T1]{fontenc}    
\usepackage{hyperref}       
\usepackage{url}            
\usepackage{booktabs}       
\usepackage{amsfonts}       
\usepackage{nicefrac}       
\usepackage{microtype}      
\usepackage{lipsum}		
\usepackage{graphicx}
\usepackage{natbib}
\usepackage{doi}
\usepackage{authblk}
\usepackage{makecell}
\usepackage{wrapfig}
\usepackage{breakcites}
\usepackage{adjustbox}
\usepackage{float}
\usepackage{multirow}
\usepackage{amsmath}
\usepackage{xcolor}
\usepackage{tabularx}
\graphicspath{ {./images/}}
\usepackage{epstopdf}
\usepackage{enumitem}
\usepackage{acronym}
\usepackage{graphicx}
\usepackage{amsmath}
\usepackage{amssymb}
\usepackage{booktabs}
\usepackage{acronym}
\usepackage{multirow}
\usepackage{tikz}
\usepackage{float}
\usepackage[capitalize]{cleveref}
\crefname{section}{Sec.}{Secs.}
\Crefname{section}{Section}{Sections}
\Crefname{table}{Table}{Tables}
\crefname{table}{Tab.}{Tabs.}

\title{Explainable Semantic Medical Image Segmentation with Style}

\author{
Wei~Dai\thanks{Principal corresponding author},
Siyu~Liu,
Craig~B.~Engstrom,
Shekhar~S.~Chandra
}
\affil[]{School of Information Technology and Electrical Engineering, The University of Queensland, Australia}

\begin{document}
\maketitle
\begin{abstract}

    Semantic medical image segmentation using deep learning has recently achieved high accuracy, making it appealing to clinical problems such as radiation therapy. However, the lack of high-quality semantically labelled data remains a challenge leading to model brittleness to small shifts to input data. Most works require extra data for semi-supervised learning and lack the interpretability of the boundaries of the training data distribution during training, which is essential for model deployment in clinical practice. We propose a fully supervised generative framework that can achieve generalisable segmentation with only limited labelled data by simultaneously constructing an explorable manifold during training. The proposed approach creates medical image style paired with a segmentation task driven discriminator incorporating end-to-end adversarial training. The discriminator is generalised to small domain shifts as much as permissible by the training data, and the generator automatically diversifies the training samples using a manifold of input features learnt during segmentation. All the while, the discriminator guides the manifold learning by supervising the semantic content and fine-grained features separately during the image diversification. After training, visualisation of the learnt manifold from the generator is available to interpret the model limits. Experiments on a fully semantic, publicly available pelvis dataset demonstrated that our method is more generalisable to shifts than other state-of-the-art methods while being more explainable using an explorable manifold.
\end{abstract}

\section{Introduction}
\label{sec:intro}

\ac{CNN}~\cite{ronneberger2015u, cciccek20163d, isensee2021nnu, isensee2017brain, dai2022can3d} have been widely used to enhance the segmentation performance for medical images. However, large quantities of training samples are often required by \ac{CNN} models to avoid over-fitting, which is especially challenging in medical image segmentation, as data scarcity is common due to the extensive time and expertise involved in labelled data acquisition. Semi-supervised and unsupervised generative adversarial methods~\cite{li2017brain,xue2018segan, yuan2019unified,rezaei2019conditional,gaj2020automated,nie2020adversarial,jafari2019semi,dong2018unsupervised,ning2018automated,xing2020lesion,decourt2020semi,lei2020skin} have been proposed to address these problems through direct translation between images and masks. While using unlabelled or multi-model data, these methods are still vulnerable to distribution shifts due to the lack of sample diversities. Mixup~\cite{zhang2017mixup} was proposed recently to diversify training samples for deep learning models through linear interpolation of random inputs. However, it still requires human intervention to balance the trade-off between model accuracy and generalizability.

To make segmentation models generalisable to distribution shifts, another group of \ac{GAN}-based methods~\cite{jin2018ct, chaitanya2019semi, shen2019learning, shin2018medical, huo2018synseg} have also been proposed to increase the diversity of training samples through augmentation. However, these methods often require a separated segmentation model trained individually with the augmented data and the segmentation training is not directly linked to adversarial learning. While some methods~\cite{yuan2019unified, zhao2018craniomaxillofacial} achieve end-to-end training with both segmentation and augmentation pipelines, they often involve multiple pairs of \acp{GAN}, which substantially increase the training complexity. Furthermore, none of the mentioned methods are explainable as they lack the style to map the input distribution, hence, it is impossible to build a manifold of the learnt features for model evaluation.

Accurate medical image segmentation is crucial for clinical diagnosis. Hence, it is critical to identify the boundary of valid input distribution for the trained model, as deep learning models tend to be unpredictable when handling inputs outside the training distribution. Confidence learning~\cite{decourt2020semi, nie2020adversarial}, which generates pixel-wise probability maps from discriminator, were proposed to validate the segmentation outputs. However, they cannot identify the learnt distribution due to the lack of style. Style-based methods~\cite{abdal2021labels4free, li2021semantic, lee2022adas, chai2022semi, shi2020novel} were proposed to map the input distribution using style by extracting multi-scale features during segmentation. However, they were neither designed to record the learnt distribution nor to visualise them for clinical evaluation. Moreover, none of them explores the possibility of StyleGAN for supervised training with limited medical data only that are semantically labelled.
\begin{figure}[!t]
    \centering
    \includegraphics[width=0.6\textwidth]{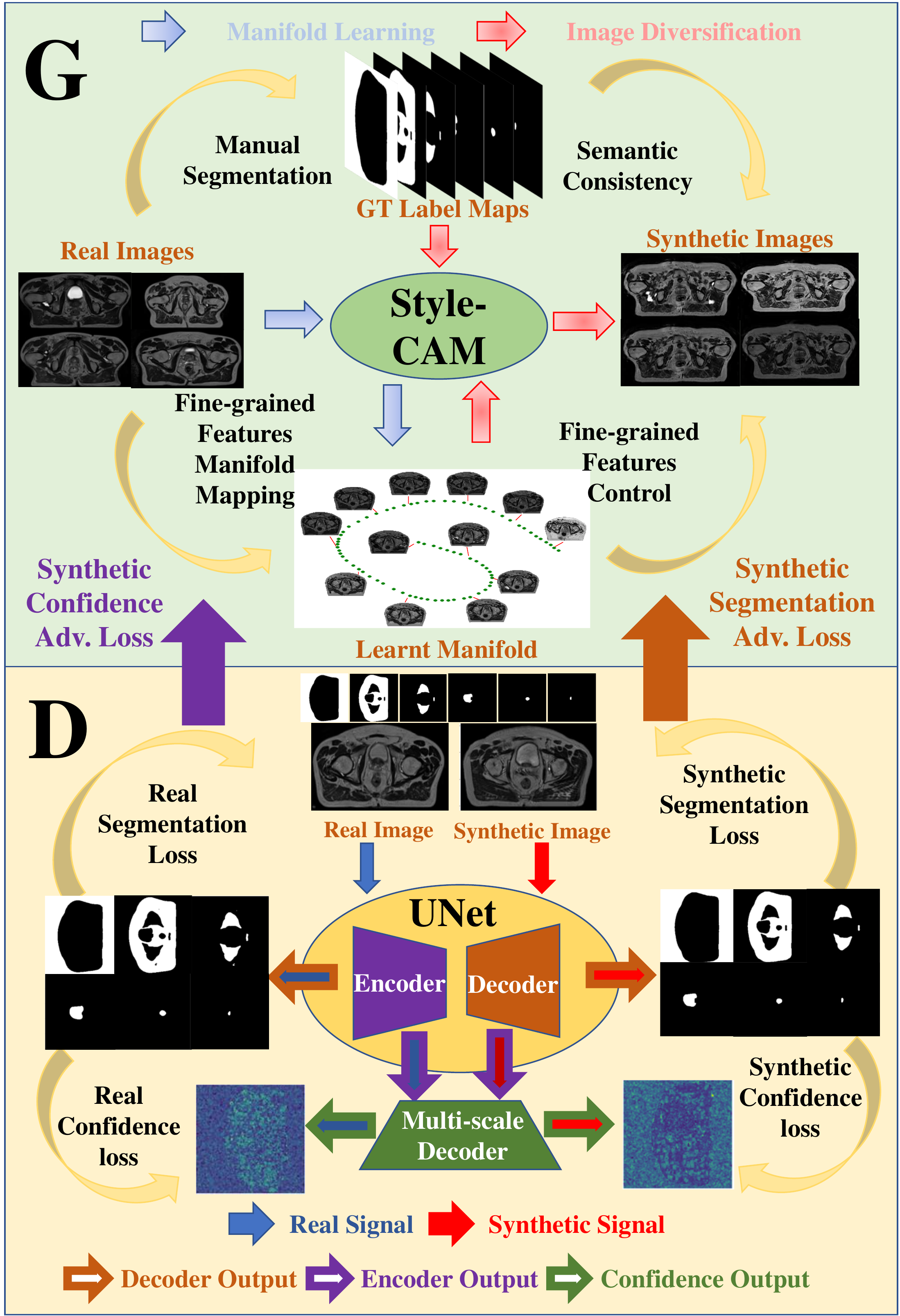}
    \caption{Our GASE framework. Generator G learns an explorable manifold using Style-CAM, which implements style-based modulated convolution with kernel dilation and synthesizes images using the learnt manifold. Discriminator D segments input images semantically while supervising the manifold learning using confidence maps and segmentation outputs.}
    \label{fig:overview}
\end{figure}
In this work, we propose \ac{GASE}, an end-to-end style-based generative model for the semantic segmentation of medical images supervised with limited labelled data. \ac{GASE} can build a manifold of the learnt input distribution entirely based on the existing labelled data during segmentation training. Instead of augmenting images through the direct translation using training samples, \ac{GASE} can diversify the samples through the combination of \ac{GT} masks and extracted features from the learnt manifold. Our aim is to improve the segmentation performance using gradients from both real and diversified samples during segmentation training. To do so, we integrate manifold-based synthesis and semantic segmentation into a generator and a discriminator respectively, while training them in an end-to-end adversarial manner. The generator utilises \ac{ModConv} with style~\cite{karras2020analyzing} as a mapping function to build the manifold of the input feature space for synthesis, and the discriminator segments input images with the help of the synthetic samples. Unlike a similar work~\cite{shi2020novel} which is based on adaptive instance normalization~\cite{karras2019style}, we introduce Style-CAM, which implements the style-based \ac{ModConv} with kernel dilation for \ac{CAM}~\cite{dai2022can3d} in our generator, to achieve multi-scale fine-grained features control without any resampling operations. Moreover, the segmentation training process requires no artificial label maps using semantic consistency, thus, preserving the target structural information within the synthetic labelled pairs, which is essential for clinical requirements. The main contributions of this paper are as follows:
\begin{enumerate}

\item We propose \ac{GASE}, a style-based generative model that improves the generalizability of the segmentation model to data shifts by enhancing the diversity of training samples via manifold learning. To the best of our knowledge, \ac{GASE} is the first method to introduce manifold learning with modulated convolution to supervised semantic medical image segmentation.

\item The proposed model achieves end-to-end training with dual functional networks, a generator that visualises the manifold of learnt distribution for model interpretation, and a discriminator that segments the medical images semantically. We also introduce style-CAM, which implements modulated convolution with kernel dilation for the first time, as far as we know.

\item We demonstrate the effectiveness of \ac{GASE} with a series of experiments on an open-access labelled pelvis MR dataset with multiple classes. \ac{GASE} outperforms previous state-of-the-art methods for the chosen dataset, especially for imbalanced classes, and demonstrates better generalisation to input distribution shifts.

\end{enumerate}

\begin{figure*}[!t]
    \centering
    \includegraphics[width=1.0\textwidth]{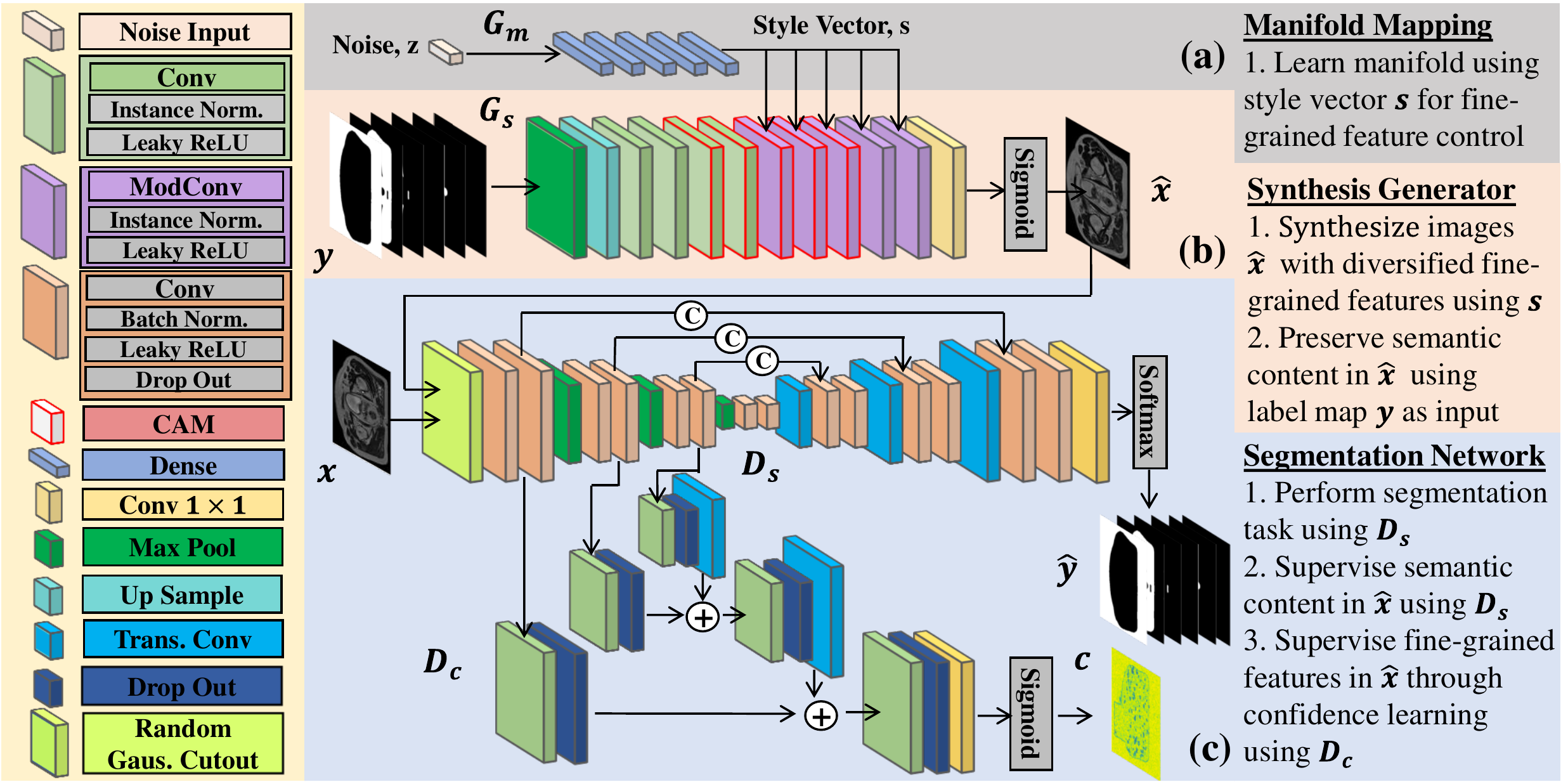}
    \caption{Overall architecture of the proposed GASE. The generator contains two modules, (a) Manifold Mapping $G_m$ and (b) Synthesis Generator $G_s$. It is used to learn the manifold of input features while diversifying training samples for segmentation based on the learnt manifold. The discriminator consists of two branches, (c) Segmentation branch $D_s$ and (d) Confidence Learning branch $D_c$. It performs semantic segmentation while guiding manifold learning. $D_s$ supervises semantic content in $\hat{x}$ and $D_c$ supervises fine features in $\hat{x}$.}
    \label{fig:overall_architecture}
\end{figure*}

\section{Related Work}

\subsection{GAN-based Segmentation Methods}
\textbf{Direct Translation Methods}: Since image segmentation can also be considered as image translation between the image domain and the label domain, \ac{GAN}-based methods, which are noted for high-quality cross-domain image translation, have also been proposed to translate medical images into label maps directly using generator~\cite{li2017brain, rezaei2019conditional, dong2018unsupervised}. Segmentation losses for the generator and multi-scale feature losses for the discriminator were added in ~\cite{zhao2018craniomaxillofacial, xue2018segan, gaj2020automated} for extra supervision during training. Multistream translation and segmentation methods \cite{yuan2019unified, xing2020lesion} based on CycleGAN\cite{zhu2017unpaired} were also proposed to utilize multi-model data under cycle consistency. Confidence learning~\cite{nie2020adversarial, decourt2020semi, jafari2019semi}, which outputs a pixel-wise probability map from the discriminator, was introduced to supervise the translation on a pixel level. Dual discriminators~\cite{lei2020skin} were also proposed to supervise the segmentation using both semantic information and contextual environment. However, these methods are vulnerable to model over-fitting due to data scarcity.

\textbf{Augmentation Methods}: To address the problem of data scarcity and make segmentation model generalizable to data shifts, \ac{GAN}-based methods were also proposed to augment the training samples through image synthesis~\cite{jin2018ct, shen2019learning}. Chaitanya et.al. \cite{chaitanya2019semi} used two sub-networks with shared intermediate layers to achieve separate control on the deformation and intensity of augmented images. Artificial label maps were also used to generate synthetic pairs for augmentation by translating them to the target image domain~\cite{shin2018medical, shi2020novel}. CycleGAN-based approach~\cite{huo2018synseg} augments synthetic pairs to train unlabelled data using labelled data from another domain. However, none of the augmentation methods model the input data distribution, hence, lacking generalization capability to distribution shifts.

\textbf{Style-based Methods}: To make segmentation more generalizable to distribution shifts, StyleGANs~\cite{karras2019style,karras2020analyzing}, which learn input distribution by extracting multi-scale features using style, has been recently proposed for segmentation tasks in computer vision ~\cite{abdal2021labels4free, li2021semantic, lee2022adas}. Abdal et.al.~\cite{abdal2021labels4free} proposed a style-based unsupervised segmentation network that is trained using synthetic images from two pre-trained generators. Li et.al.~\cite{li2021semantic} tackled semantic segmentation using StyleGAN by modelling the joint image-label distribution for unlabeled images using limited labelled ones in a semi-supervised manner. Chai et.al.~\cite{chai2022semi} also proposed a semi-supervised method to segment medical images by generating attention-based label maps during a two-stage hierarchical synthesis using style-based generators.

\section{Methodology}
\subsection{Definition and Overview}
\label{sec:def_over}



We denote the limited labelled dataset as $S = \{x, y\}$, where $x \in \mathbb{R}^{B \times H \times W \times 1}$ represents a batch of images under real image distribution $P_{data}(x)$ with batch size $B$, $y \in \mathbb{R}^{B \times H \times W \times C}$ is the corresponding one-hot encoded \ac{GT} label batch with $C$ classes. The overall architecture of \ac{GASE} is shown in \cref{fig:overall_architecture}. The generator $G$ contains two sub-models: manifold mapping $G_m: z \rightarrow s$, which generates style vectors $s \in \mathbb{R}^{1 \times P}$ given noise vectors $z \in \mathbb{R}^{1 \times L}$, and the synthesis generator $G_s : (y, s) \rightarrow \hat{x}$, where $\hat{x}$ represents synthesized images from a recovered fake distribution $P_{fake}(x)$ after manifold learning. The discriminator $D: (x, \hat{x}) \rightarrow (\hat{y}_x, \hat{y}_{\hat{x}}, c_x, c_{\hat{x}})$ is a segmentation network which takes both $x$ and $\hat{x}$ as input and outputs predicted label map $\hat{y} \in \mathbb{R}^{B \times H \times W \times C}$ and confidence map $c \in \mathbb{R}^{B \times H \times W \times 1}$ for both inputs simultaneously.

Overall, we propose a style-based $G$ that controls the features within $\hat{x}$ by injecting learnable $s$ into the intermediate layers of $G_s$ using \ac{ModConv}. As $G$ gets better at synthesizing $\hat{x}$, different $s$ will be mapped to a manifold that recovers the distribution of input features. By using $y$ as input for $G$ and outputting $\hat{y}$ from $D$, semantic consistency can be achieved during adversarial training so the semantic content from input $y$ will be preserved in the corresponding $\hat{x}$. Such consistency will make manifold mapping focus more on fine-grained features, such as pixel density and contrast, and preserve global features, such as real semantic information from $y$. The whole manifold mapping process is supervised by the pixel-wise probability map $c$ from $D$ through adversarial training. By feeding different $s$, extracted from the learnt manifold, with the same $y$ into $G$, images with the same semantic content as $y$ but diversified fine-grained features could be synthesized, thus, forming a new synthetic pair $(\hat{x}, y)$ that shares the same label map with the corresponding real pair $(x,y)$. During segmentation training, gradient information from the $(\hat{x}, y)$ is progressively propagated along with the $(x,y)$ as the $P_{fake}(x)$ better resembles the $P_{data}(x)$. Once the training is done, $G$ can be used to visualize the learnt manifold by projecting $s$ into the lower dimension with the corresponding $\hat{x}$. We now discuss the technical details for each part of the proposed architecture.

\subsection{Analyszing Role of Style-based Generator}
Two strategies are taken to make the segmentation task more generalizable to distribution shifts under limited training samples. First, we want to diversify the training samples while preserving the semantic content from \ac{GT} label maps. Doing so will avoid using artificial label maps when generating synthetic labelled pairs, preventing augmentation leakage~\cite{karras2020training}. Second, we want to make the model interpretable so the training data distribution can be mapped into a manifold for visualization. Based on these ideas, we proposed a style-based generator with two sub-models, manifold mapping $G_m$ and synthesis generator $G_s$

\textbf{Manifold Mapping, $G_m$}: As shown in \cref{fig:overall_architecture}a, $G_m$ is designed to generate style vectors $s$ that control the fine-grained features within the $\hat{x}$. Given a Gaussian noise vector $z$, $G_m$ will transform $z$ into style vector $s$ using consecutive dense layers. Each $s$ is learned during training to match a specific fine-grained feature from the input distribution. However, due to the limited training samples, over-fitting could easily occur when all $z$ are transformed into the same $s$. To address this, we propose a penalizing term when training $G$ by regularizing the transformation as follows:
\begin{equation}
\label{eqn:gpen_loss} 
    \mathcal{L}_{Gp}(y,z_1, z_2 ; \theta_{G_m, G_s}) =\frac{\lambda_1}{ \mathcal{L}_{mse}( \hat{x}_1,  \hat{x}_2) + \lambda_2},
\end{equation}
where $\hat{x}_1$ and $\hat{x}_2$ are two synthetic images generated using different noise input $z_1$ and $z_2$. Taking the reciprocal of \ac{MSE} between two synthetic images, a large loss will be applied when similar images are generated with different $z$. By doing so, $s$ will be forced to learn the correlation among different input features, building a manifold with diversified features that ensemble $P_{data}(x)$ as well as possible. Values of 0.01 and 0.001 are chosen for $\lambda_1$ and $\lambda_2$, to limit the maximum penalty with a sharp gradient. Moreover, to avoid sparse distribution of $s$ and distribute them evenly across the manifold, spectral normalization~\cite{miyato2018spectral} is applied to each dense layer in $G_m$ to mitigate large gradient of $\theta_{G_m}$ during training.

\begin{figure}[!t]
    \centering
    \includegraphics[width=0.6\textwidth]{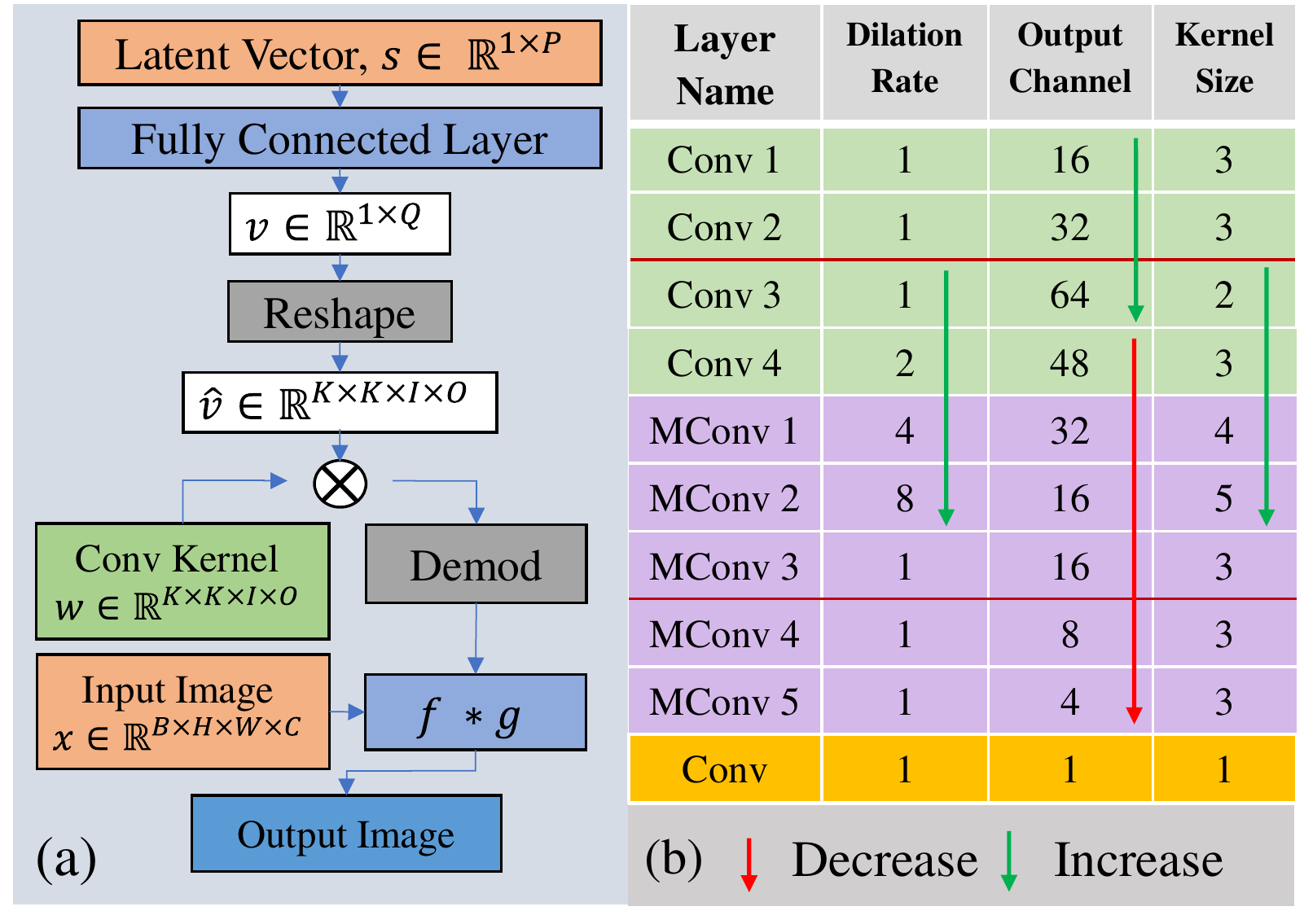}
    \caption{a) Structure of the modified modulated convolution layer. b) Details for convolution layers within synthesis generator $G_m$. Layers between two red lines represent the layers within the context aggregation module (CAM).}
    \label{fig:modconv_architecture}
\end{figure}

\textbf{Modulated Convolution Layer}: To map $s$ from $G_m$ to various fine-grained features within input images through manifold learning, a modified \ac{ModConv} layer shown in \cref{fig:modconv_architecture}a is designed to modulate convolution kernel using $s$. Unlike the original \ac{ModConv} layer~\cite{karras2020analyzing}, which uses different $s$ for different layers during multi-scale features extraction, the same $s$ is used instead to match a unique feature during manifold learning. To achieve this, a dense operation is firstly applied to transform $s \in \mathbb{R}^{1 \times P}$ to $v \in \mathbb{R}^{1 \times Q}$, so the vector size $Q$ matches the number of trainable weights within convolution kernel $w \in \mathbb{R}^{K \times K \times I \times O}$, where K is the kernel size, I and O represent layer input and output channel numbers respectively. After reshaping $v$, $\hat{v} \in \mathbb{R}^{K \times K \times I \times O}$ is used to modulate convolution kernel through simple multiplication: $\hat{w} = w \cdot \hat{v}$. Finally, the same demodulation procedure in \cite{karras2020analyzing} is implemented to normalize the scaled kernel weights, and the resulting kernel is used to conduct convolution operation with layer input.

\textbf{Synthesis Generator, $G_s$}: As illustrated in \cref{fig:overall_architecture}b, multiple standard convolution blocks are first applied to learn the general shape information of the input masks. Then a modified version of \ac{CAM}~\cite{dai2022can3d} is implemented with \ac{ModConv} layers to modify the fine-grained features on multiple scales under a larger receptive field without any re-sampling operations. As shown in \cref{fig:modconv_architecture}b, different to the original \ac{CAM}, the kernel sizes of the convolution layers are increased gradually to expand the regions that could be modulated by $s$. Concurrently, the number of output channels for each layer is progressively decreased to avoid model over-parameterization. In this supervised learning, other than the penalty term introduced in \cref{eqn:gpen_loss}, all other training signals for $G_s$ only come from the $D$. By taking the same $y$ and different $s$, $G_s$ can diversify the training samples with various fine-grained features while preserving the semantic content from $y$.

\subsection{Segmentation Discriminator with Supervision}
As illustrated in \cref{fig:overall_architecture}c, the proposed $D$ consists of two branches, a segmentation branch $D_s$ and a confidence learning branch $D_c$. $D_s$ is the backbone of \ac{GASE} that performs image segmentation tasks. At the same time, it also supervises semantic content of $\hat{x}$ during $G$ training. $D_c$ is designed to supervise the fine-grained features within the $\hat{x}$ through confidence learning.

\textbf{Discriminator for Segmentation}: The $D_s$ is designed based on the 2D-UNet~\cite{ronneberger2015u} with additional dropout for image segmentation. By training $D_s$ with the synthetic pair $(\hat{x},y)$ and the real pair $(x,y)$ simultaneously, the different fine-grained features with the same label map between $x$ and $\hat{x}$ will force $D_s$ to handle more diversified features during segmentation, hence, becoming more robust to distribution shifts. The loss for such counter pairs is defined as:
\begin{equation}
\label{eqn:dseg_loss}
    \mathcal{L}_{Ds}(x, \hat{x} :\theta_{D_s}) = \\
    \mathcal{L}_{dsl}(D_s(x), y) + \lambda_3 \cdot \gamma \cdot \mathcal{L}_{dsl}(D_s(\hat{x}), y),
\end{equation}
where $\mathcal{L}_{dsl}$ is the weighted \ac{DSL} proposed in \cite{dai2022can3d} and $\lambda_3$ is a weighting factor which limits the maximum signal provided by the $\hat{x}$. To limit the second loss term when the quality of $\hat{x}$ is low at the initial training stage, a scheduled weighting factor $\gamma$, which starts from zero, is increased progressively based on the inverse of polynomial decay until reaching one when $G$ is capable of synthesizing realistic $\hat{x}$. Moreover, to give $G$ more freedom during synthesis and balance the training between $G$ and $D$, we designed a random cutout layer with Gaussian noise to inject local randomness into the input of $D$ ( \cref{fig:overall_architecture}c). Compared to the original random cutout~\cite{devries2017improved}, which uses the constant value in the cutout region, Gaussian noise is used instead for better gradient propagation.

\begin{figure}[!t]
    \centering
    \includegraphics[width=0.7\textwidth]{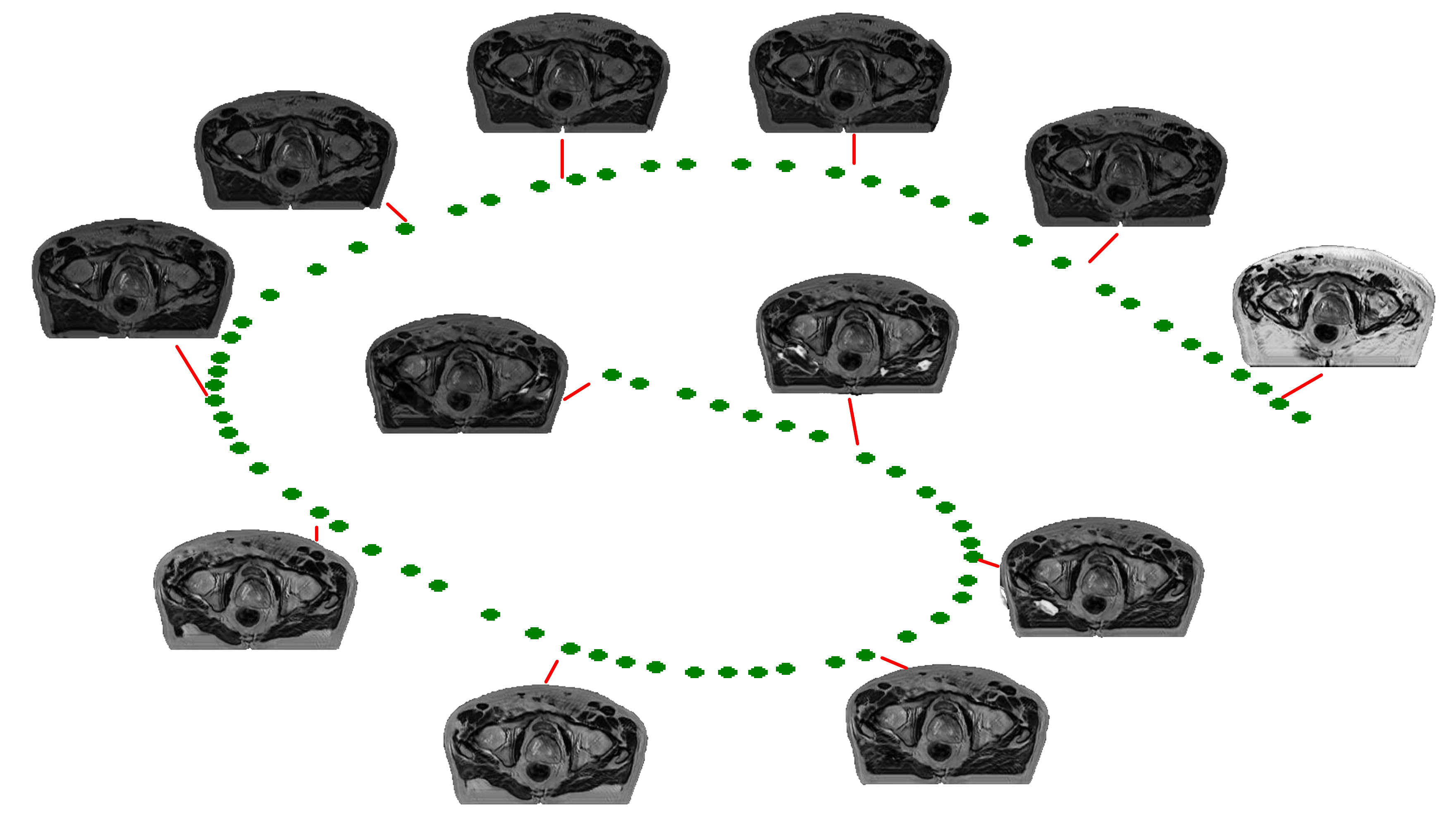}
    \caption{Visualization of the manifold learnt by \ac{GASE} during segmentation training. Green dots represent the sampled style vectors projected onto a 2D plane. Each style vector represents a specific fine-grained feature shown in the provided images. Such a manifold can be used to explain the segmentation training.}
    \label{fig:manifold_visual}
\end{figure}

\textbf{Discriminator for Supervision}: While getting better at the segmentation task, $D_s$ also supervises the semantic content within $\hat{x}$ by propagating segmentation loss between $y$ and $\hat{y}$ to $G$ during adversarial training, which is defined as:
\begin{equation}
\label{eqn:gseg_loss}
    \mathcal{L}_{Gs}(y,z ; \theta_{G_m, G_s}) = \mathcal{L}_{dsl}(D_s(\hat{x}), y).
\end{equation}
By minimizing $\mathcal{L}_{Gs}$, $G$ is forced to synthesize images with the same semantic content as $y$, therefore, achieving semantic consistency between $\hat{x}$ and $y$. To supervise fine-grained features within $\hat{x}$, the encoder of $D_s$ also extracts features from the input images for confidence branch $D_c$. Compared to the standalone encoder proposed in \cite{xue2018segan}, the features extracted in our approach are constrained by the multi-class shape information from $y$ as the encoder is trained under the guidance of the decoder, which performs shape reconstruction. Such shape-constrained features will provide class information to $D_c$ when generating confidence map $c$.

As illustrated in \cref{fig:overall_architecture}c, $D_c$ uses multi-scale features extracted by the encoder of $D_s$ as input. It supervises $G$ by providing an adversarial loss based on the generated probability map $c \in \mathbb{R}^{B \times H \times W \times 1}$ that indicates the pixel-wise realness of fine-grained features within $\hat{x}$. Due to the down-sampling operations within the encoder of $D_s$, large-scaled features will cover more local information, whereas lower-scaled features will contain more global information. By aggregating the multi-scale features, the generated confidence map will better distinguish the ’real’ or ’fake’ features based on local and global information. The confidence loss when training $D_c$ is defined as:
\begin{equation}    
\label{eqn:dc_loss}
    \mathcal{L}_{Dc}(x,\hat{x} ; \theta_{Dc}) = \mathcal{L}_{bce}(c_{x}, (1 - \zeta)) + \mathcal{L}_{bce}(c_{\hat{x}}, \zeta),
\end{equation}
where $\mathcal{L}_{bce}$ is the binary cross-entropy loss, $c_{x}$ and $c_{\hat{x}}$ are the confidence maps generated with $x$ and $\hat{x}$ respectively, $\zeta$ is a small Gaussian noise used to provide better gradient. The adversarial confidence loss for $G$ is defined as:
\begin{equation}    
\label{eqn:gc_loss}
    \mathcal{L}_{Gc}(y,z ; \theta_{G_m, G_s}) = \mathcal{L}_{bce}(c_{\hat{x}}, (1 - \zeta)).
\end{equation}

\subsection{Training Losses Details}
During the training, $G$ and $D$ will be trained iteratively for specified epochs. We define the hybrid adversarial loss function for $G$ as:
\begin{equation}
\label{eqn:goverall_loss}
    \mathcal{L}_G\left(\theta_{G_m} , \theta_{G_s}\right) = \beta \mathcal{L}_{Gs} + \mathcal{L}_{Gc} + \mathcal{L}_{Gp},
\end{equation}
where $\beta$ is set to 2 to enforce the semantic consistency. The hybrid loss function defined for $D$ is:
\begin{equation}
\label{eqn:doverall_loss}
    \mathcal{L}_D\left(\theta_{D_s} , \theta_{D_c}\right) = \mathcal{L}_{Ds} + \mathcal{L}_{Dc}.
\end{equation}

\section{Experiments}

\begin{table*}[!t]
\centering
\setlength{\tabcolsep}{3pt}  
\caption{The ablation study for each module of the proposed \ac{GASE}. The modules are $G_s$: synthesis generator, $G_m$: manifold mapping, $D$: segmentation discriminator and Mixup: mixup augmentation technique.}

\scalebox{0.75}{

\begin{tabular}{cccc|clc|ccc|ccc}
\hline
\multirow{2}{*}{$G_s$} & \multirow{2}{*}{$G_m$} & \multirow{2}{*}{D}        & \multirow{2}{*}{Mixup} & \multicolumn{3}{c|}{DSC}                                             & \multicolumn{3}{c|}{HD}                                     & \multicolumn{3}{c}{MSD}                                      \\ \cline{5-13} 
                       &                        &                           &                        & Bladder            & \multicolumn{1}{c}{Rectum} & Prostate           & Bladder            & Rectum            & Prostate           & Bladder            & Rectum             & Prostate           \\ \hline
                       &                        & \checkmark &                        & 0.91 ± 0.07          & 0.93 ± 0.05                  & 0.72 ± 0.16          & 10.2 ± 13.9        & 4.70 ± 11.4         & \textbf{8.08 ± 5.22} & 1.76 ± 1.36          & \textbf{1.13 ± 1.67} & 3.61 ± 1.83          \\
\checkmark                      &                        & \checkmark                         &                        & 0.91 ± 0.07          & 0.93 ± 0.05                  & 0.73 ± 0.19          & 12.0 ± 22.1        & \textbf{4.64 ± 9.20} & 9.89 ± 19.1         & 1.75 ± 2.14          & 1.16 ± 1.69          & 4.28 ± 14.9         \\
                       &                        & \checkmark                         & \checkmark                      & 0.92 ± 0.07          & 0.92 ± 0.07                  & 0.74 ± 0.16          & 7.78 ± 15.2          & 4.73 ± 5.19         & 8.19 ± 4.22          & 1.44 ± 0.93          & 1.25 ± 1.17          & 3.46 ± 1.66          \\
\checkmark                      & \checkmark                      & \checkmark                         &                        & \textbf{0.92 ± 0.06} & \textbf{0.93 ± 0.02}         & \textbf{0.75 ± 0.17} & \textbf{6.67 ± 7.79} & 5.14 ± 8.75         & 8.52 ± 6.09          & \textbf{1.37 ± 0.83} & 1.21 ± 1.23          & \textbf{3.31 ± 1.91} \\ \hline
\end{tabular}}

\label{tab:ablation_GASE_parts}
\end{table*}

\textbf{Datasets}:
Due to the lack of public medical datasets that are fully segmented semantically with multiple classes (including body and bone), an open-access \ac{MR} dataset of the pelvis~\cite{dowling2015automatic} was used for our experiment. The 3D pelvis dataset consists of 211 \ac{MR} examinations acquired each week longitudinally for up to 8 weeks for a cohort of 38 patients diagnosed with prostate cancer. Single 3D CT image was also acquired at baseline for each patient and used for cross-modality evaluation. All images are manually segmented into six classes: background, pelvic cavity, bone, bladder, prostate and rectum. Each examination has an image size of $256 \times 256 \times 256$ voxels with a resolution of $1.68 \times 1.68 \times 1.56$ mm.

\textbf{Preprocessing}:
To make the 3D images suitable for our 2D models, 2D slices containing all classes were extracted from each volume along the axial axis after bias-field correction~\cite{tustison2010n4itk}. The resulting 1775 slices were then used as our training set after being normalised to $[0, 1]$ in intensity values with zero mean and standard deviation. Finally, three-fold cross-validation studies were conducted by randomly splitting the extracted slices into three subsets (751/656/368) on a patient basis. The testing set from each fold was also used as the validation set during training. All results reported in this paper are the averaged 3-fold cross-validation results using the testing set only.

\textbf{Implementation}:
We employed the proposed model with the Tensorflow v2.9.0 framework on NVIDIA Volta V100 GPU with 32GB of memory. All models were trained for 2000 epochs with a batch size of 15. The model with the best validation performance during training was saved for evaluation. Adam optimizers with initial learning rates of $5 \times 10^{-3}$ and $1 \times 10^{-3}$ were chosen for the training of $G$ and $D$ respectively. Furthermore, the polynomial decay was used to decrease both learning rates with a power of two after each epoch until reaching zero at the end of training.

\textbf{Evaluation Metrics}:
Three evaluation metrics were adopted in this experiment. The \ac{DSC} was used to access the overlapping ratios between the $\hat{y}$ and $y$. \ac{HD} and \ac{MSD} values were used to calculate the maximum and averaged surface distance error between the anatomical boundaries of the $\hat{y}$ and $y$.

\subsection{Ablation Studies}

\cref{fig:manifold_visual} visualises the manifold of fine-grained features of the inputs learnt by $G$ during adversarial training. To visualise such a manifold, noise vectors, $z$, were randomly chosen to generate different style vectors $s$ using $G_m$. The manifold was then obtained by projecting the generated $s$ onto a 2D plane using T-SNE~\cite{van2008visualizing}. The corresponding synthetic images $\hat{x}$ used for manifold visualisation were then generated by combining different $s$ with the same label input $y$ using $G_s$ and were placed in the 2D plane based on the location of their corresponding $s$. As demonstrated in \cref{fig:manifold_visual}, all synthetic images have the same semantic content with fine-grained features, such as pixel density and contrast, gradually changing along the learnt manifold. This indicates that the proposed model successfully explores the possible combinations of fine-grained features extracted from the limited dataset, which are then used to improve the segmentation performance by diversifying the training samples based on the learnt manifold. Such visualisation could also be used to explain the trained model by displaying all valid inputs that $ D$ has seen during segmentation training.

\cref{tab:ablation_GASE_parts} gives the ablation study of each module in our approach. To evaluate the baseline performance of the segmentation model $D_s$, we trained it using the limited labelled data both with and without mixup~\cite{zhang2017mixup} augmentation. As shown in \cref{tab:ablation_GASE_parts}, the diversified training samples after mixup improve the performance of the $D_s$ as expected. To evaluate the performance of our model with and without the \ac{ModConv} layer using $s$ from $G_m$, we separated $G_m$ from $G_s$ by replacing the \ac{ModConv} layer with the standard convolution. As shown in the table, training $D$ with $G_s$ makes the segmentation performance worse than the $D$ without it. Such results are expected as the synthesized images lose the fine-grained feature control due to the missing manifold learnt by $G_m$. However, when we combined $G_s$ with $G_m$ using \ac{ModConv} during $D$ training, the segmentation performance increases substantially and even surpasses $D$ with mixup for both \ac{DSC} and \ac{MSD}. Such results indicate that our method successfully improves the segmentation performance by diversifying the training samples based on the learnt manifold.

\begin{table*}[!t]
\centering
\caption{Comparison of quantitative results for \ac{GASE} in semantic segmentation averaged cross-validation testing results.}
\scalebox{0.8}{
\begin{tabular}{ccccccc}
\hline
      Region &  
      Mean Metric &
      GASE &
      \begin{tabular}[c]{@{}c@{}} nnUnet \\ \cite{isensee2021nnu}\end{tabular} & 
      \begin{tabular}[c]{@{}c@{}} ConfiGAN \\ \cite{nie2020adversarial}\end{tabular} & 
      \begin{tabular}[c]{@{}c@{}} IUNET \\ \cite{isensee2017brain}\end{tabular} &
      \begin{tabular}[c]{@{}c@{}} UNET \\ \cite{ronneberger2015u}\end{tabular} 
 \\ \hline

\multirow{3}{*}{Rectum}   & DSC         & \textbf{0.94 ± 0.04} & 0.92 ± 0.06          & 0.88 ± 0.09 & 0.87 ± 0.10   & 0.91 ± 0.07 \\
                          & HD(mm)      & \textbf{4.29 ± 6.87} & 5.23 ± 11.9          & 7.41 ± 7.09 & 13.8 ± 25.0  & 9.17 ± 19.3 \\
                          & MSD(mm)     & \textbf{1.11 ± 0.98} & 1.63 ± 4.75          & 1.93 ± 1.78 & 2.68 ± 3.48  & 1.96 ± 3.00  \\ \hline
\multirow{3}{*}{Prostate} & DSC         & \textbf{0.78 ± 0.16} & 0.76 ± 0.16          & 0.74 ± 0.18 & 0.71 ± 0.17  & 0.71 ± 0.23 \\
                          & HD(mm)      & 8.21 ± 9.33          & \textbf{7.37 ± 7.08} & 10.5 ± 17.4 & 16.1 ± 32.0  & 23.7 ± 62.5 \\
                          & MSD(mm)     & \textbf{3.30 ± 6.01} & 3.39 ± 5.20          & 4.78 ± 15.3 & 5.64 ± 16.5  & 18.2 ± 60.6 \\ \hline
\multirow{3}{*}{Bladder}  & DSC         & \textbf{0.90 ± 0.10}  & 0.90 ± 0.11           & 0.82 ± 0.17 & 0.85 ± 0.17  & 0.85 ± 0.17 \\
                          & HD(mm)      & 8.43 ± 9.95          & \textbf{8.34 ± 14.9} & 55.7 ± 28.5 & 20.5 ± 32.6  & 15.7 ± 27.5 \\
                          & MSD(mm)     & \textbf{1.71 ± 1.66} & 2.16 ± 6.36          & 9.26 ± 16.6 & 4.57 ± 13.0 & 4.47 ± 19.9 \\ \hline
                          
\multirow{3}{*}{Bone}     & DSC         & \textbf{0.93 ± 0.02} & \textbf{0.93 ± 0.02} & 0.90 ± 0.03  & 0.90 ± 0.04   & 0.91 ± 0.03 \\
                          & HD(mm)      & 21.1 ± 14.8          & \textbf{15.9 ± 12.7} & 41.8 ± 17.7 & 28.7 ± 17.8  & 25.9 ± 17.6 \\
                          & MSD(mm)     & 2.08 ± 0.80          & \textbf{1.84 ± 0.56} & 3.95 ± 1.86 & 3.02 ± 1.41  & 2.51 ± 1.33 \\ \hline

\end{tabular}}

\label{tab:model_comparison}
\end{table*}

\begin{figure*}[!t]
    \centering
    \includegraphics[width=1.0\textwidth]{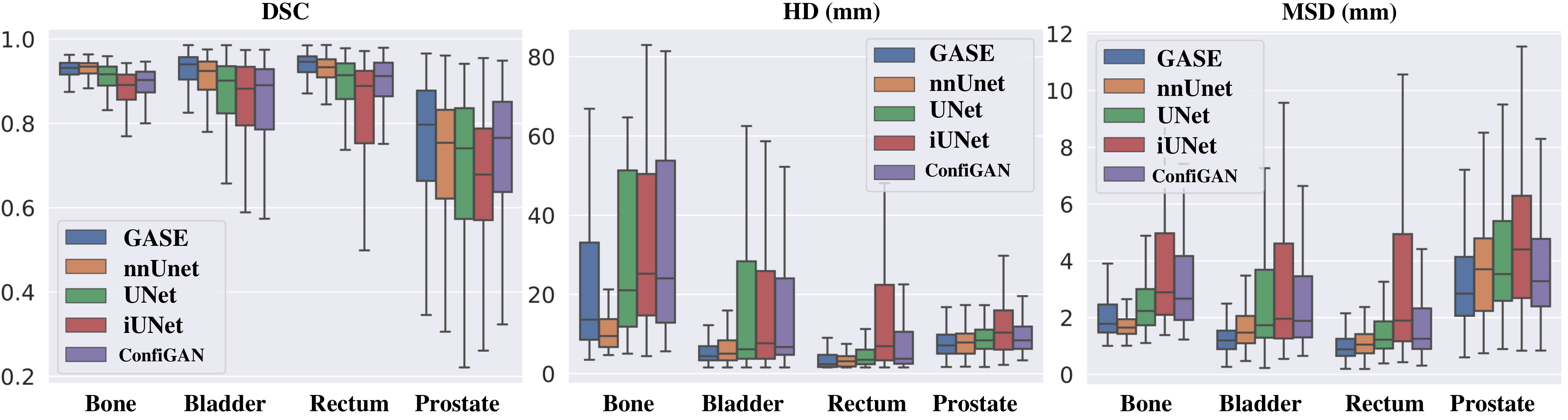}
    \caption{Testing results of \ac{GASE} and the chosen models for the validation fold containing extreme data shifts within its testing set due to extreme abdominal body fat and non-normal bladder contrast. The outliers are removed from the plot.}
    \label{fig:model_boxplot_visual_set_3}
\end{figure*}

\begin{figure*}[!t]
    \centering
    \includegraphics[width=1.0\textwidth]{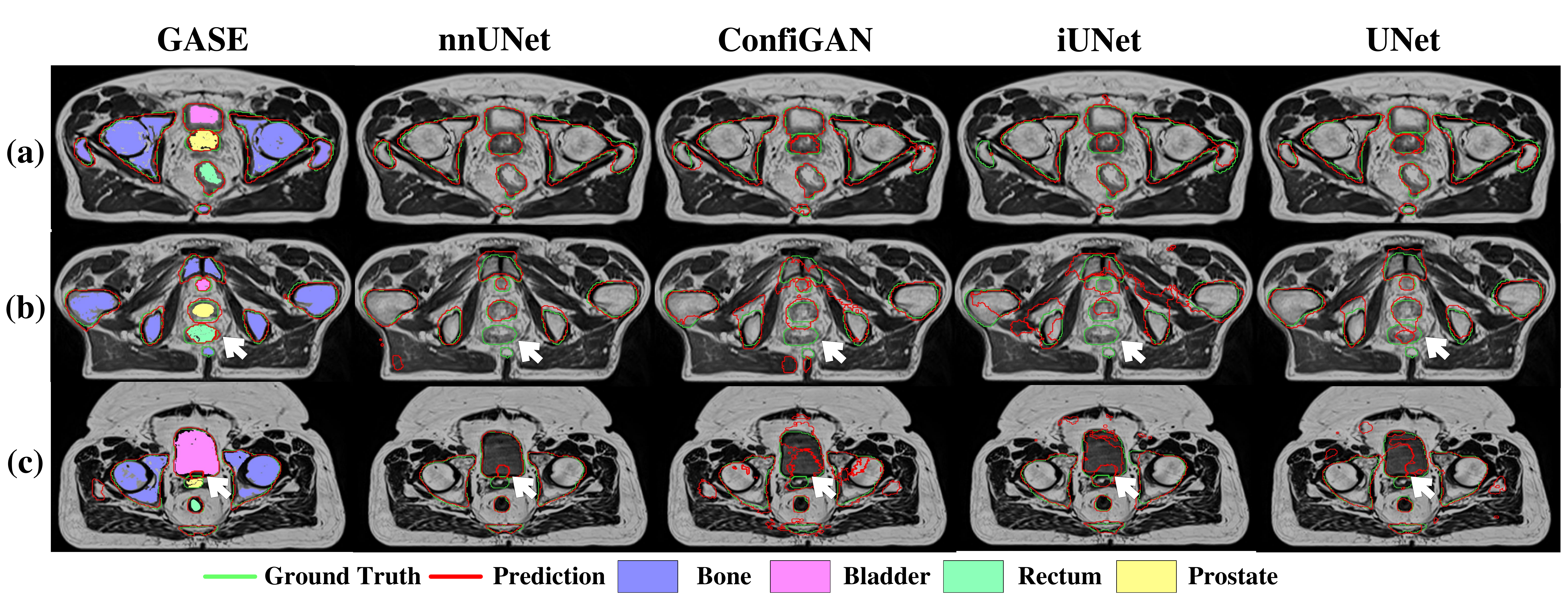}
    \caption{Illustration of segmentation results for three cases. a) Easy case when all classes have distinct features. b) Difficult case when all classes have intricate features, especially for the imbalanced classes like the rectum and prostate. c) Data shifting case when the patient has a large amount of abdominal fat and abnormal bladder contrast. Segmentation outlines are drawn in red (Prediction) and green (GT).}
    \label{fig:seg_visual}
\end{figure*}

\subsection{Comparison with State-of-the-Art Methods}

State-of-the-art supervised methods were chosen to evaluate the performance of our \ac{GASE} with limited labelled data only. Popular UNet~\cite{ronneberger2015u} and improved UNet~\cite{isensee2017brain} were chosen as the baseline models for comparison. \ac{GASE} was also compared with nnUNet~\cite{isensee2021nnu} directly, which is an automatic configuration pipeline based on UNet. \ac{GAN}-based model with confidence learning~\cite{nie2020adversarial} was chosen as the related generative model for comparison. 

\cref{tab:model_comparison} shows the quantitative results for the chosen models. \ac{GASE} performs the best in terms of \ac{DSC} and \ac{HD} for non-rigid components such as the bladder, prostate and rectum segmentation. It performs slightly worse than nnUnet on \ac{HD} and does not perform as well as nnUnet for the rigid component such as bone. The reason is that
fine-grained features of the bone are relatively consistent across all training samples, which limits the features that could be diversified during image synthesis, hence, the segmentation accuracy of the bone. The visualisation of segmentation results is also demonstrated in \cref{fig:seg_visual}. As shown in \cref{fig:seg_visual}a, all models achieve accurate segmentation results when all classes have distinct features such as density, contrast and shape from each other. However, as the features among classes become more intricate, for example, the smaller bladder, multiple sections of bone, and similar features between rectum and prostate, shown in \cref{fig:seg_visual}b. \ac{GASE} is the only model that can consistently produce accurate label maps for all classes, especially the rectum, which is missing in all other models' results. Moreover, it is found that \ac{GASE} performs substantially better than other models for one validation fold whose testing set contains one case with extreme data shifts compared to the corresponding training set. The results are displayed in \cref{fig:model_boxplot_visual_set_3}. As demonstrated in \cref{fig:seg_visual}c, the extreme case has an abnormal bladder contrast with substantially more body fat compared to images in a) and b), and only \ac{GASE} and nnUnet could segment all classes properly. However, by comparing the segmented prostate from both models, \ac{GASE} successfully mitigates the misclassified prostate within the bladder. The nnUnet, on the other hand, falsely segments a large portion of the bladder as the prostate due to the unseen bladder contrast. Both quantitative and qualitative results indicate that our \ac{GASE} is more generalisable to distribution shifts than other models.

\begin{figure}[!t]
    \centering
    \includegraphics[width=0.8\textwidth]{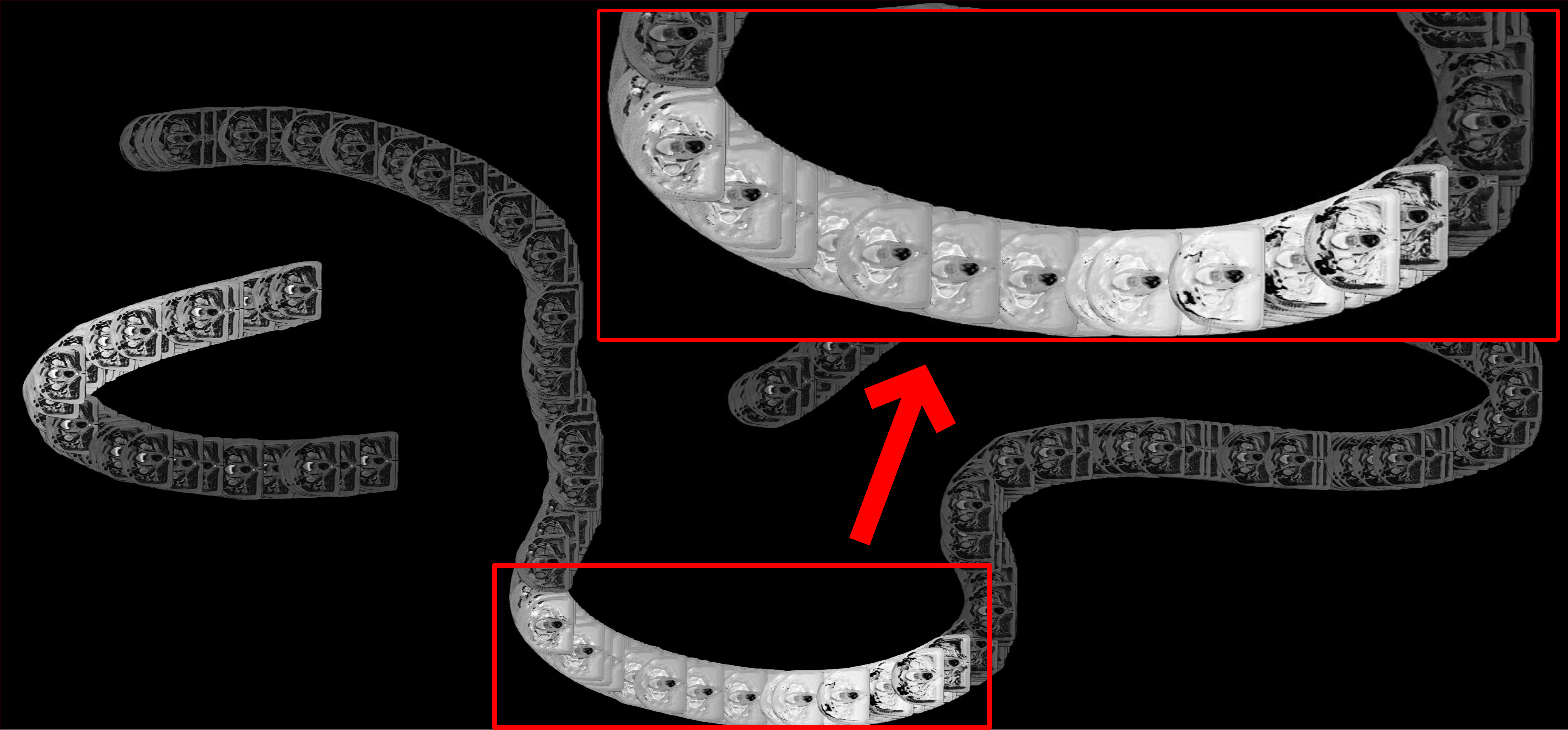}
    \caption{The visualisation of manifold learnt by \ac{GASE} when mixing both MRI and CT images as training samples. The red section shows the location of CT images within the manifold. This figure is similar to \cref{fig:manifold_visual} but with all images plotted.}
    \label{fig:manifoldCT_visual}
\end{figure}

\subsection{Generalisation to Input Diversity}
\begin{figure}[!t]
    \centering
    \includegraphics[width=0.6\textwidth]{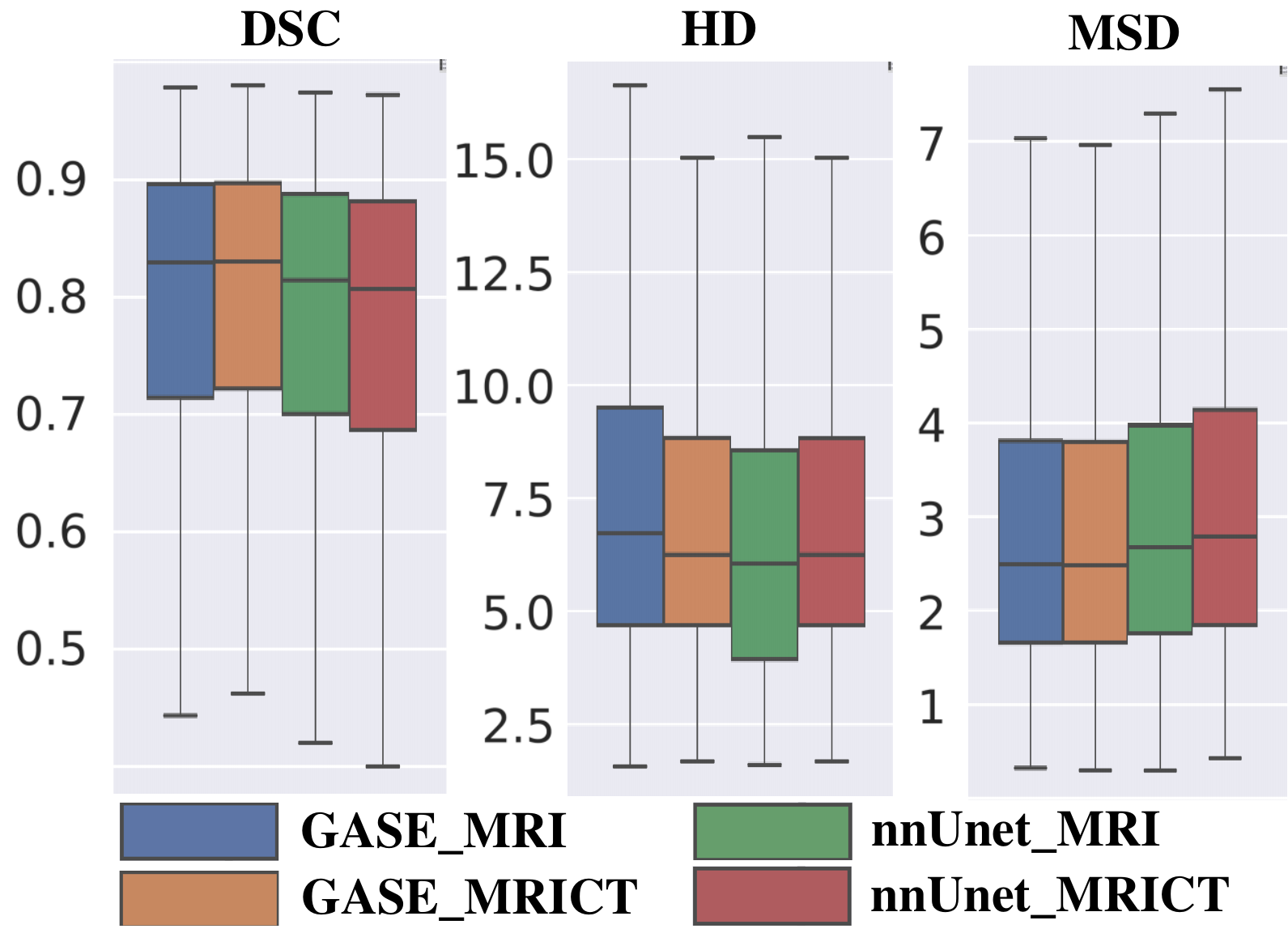}
    \caption{Box-plot results for \ac{GASE} and nnUnet when both models are trained with a mixture of MRI and CT images. With the extra CT images as input, GASE's performance increases slightly while nnUnet's performance decreases.}
    \label{fig:MRICT_box}
\end{figure}

To test how generalisable the \ac{GASE} is when training with a more complex dataset, we trained \ac{GASE} and nnUnet with both MRI and CT images provided and tested them with the original MR testing set. As shown in \cref{fig:manifoldCT_visual}, \ac{GASE} successfully distinguishes the CT from MR images by grouping images with CT features together within the learnt manifold. \cref{fig:MRICT_box} illustrates that while the extra CT data decreases the performance of nnUnet, the accuracy of \ac{GASE} increases slightly for \ac{DSC} and \ac{HD}. Therefore, unlike the nnUnet, whose performance is easily influenced by the quality of the inputs, \ac{GASE} is more generalisable to a diversified training set due to the learnt manifold, providing consistent accuracy even if the complexity of the training set is increased.

\section{Conclusion}
In this paper, we propose GASE, a style-based generative framework that achieves supervised segmentation that generalises to distribution shifts with limited labelled data. We proposed a synthesis generator and a segmentation discriminator trained in a novel way using adversarial losses incorporating confidence. Our GASE learns the manifold for segmentation with the discriminator guiding the manifold learning by supervising semantic content and fine-grained features within the synthetic samples. This manifold can be visualised to explain the trained model by demonstrating the features that the discriminator sees during segmentation training. Our GASE outperforms four state-of-the-art methods on an open-access semantic pelvis MR data set with substantially higher accuracy when handling images with extreme distribution shifts. GASE also shows consistent performance regardless of the increased complexity of the training set, but requires fully semantically labelled datasets for optimal performance.
\acrodef{CNN}{Convolutional Neural Networks}
\acrodef{GAN}{Generative Adversarial Networks}
\acrodef{GT}{ground truth}
\acrodef{DL}{Deep Learning}
\acrodef{DNN}{Deep Neural Networks}
\acrodef{MSE}{Mean Squared Error}
\acrodef{FC}{Fully Connected}
\acrodef{ModConv}{Modulated Convolution}
\acrodef{CAM}{Context Aggregation Module}
\acrodef{DSL}{Dice Squared Loss}
\acrodef{MR}{magnetic resonance}
\acrodef{DSC}{Dice Similarity Coefficient}
\acrodef{HD}{Hausdorff Distance}
\acrodef{MSD}{Mean Surface Distance}
\acrodef{ANOVA}{Analysis of Variance}
\acrodef{GASE}{Generative Adversarial Segmentation Evolution}


\bibliographystyle{unsrtnat}
\bibliography{references}

\end{document}